# Frictional characteristics of exfoliated and epitaxial graphene

Young Jun Shin [a,b], Ryan Stromberg [c], Rick Nay [c], Han Huang [d], Andrew T. S. Wee [d], Hyunsoo Yang [a,b,*], Charanjit S. Bhatia [a]

[a] Department of Electrical and Computer Engineering, National University of Singapore, 4 Engineering Drive 3, 117576, Singapore

[b] NUSNNI-NanoCore, National University of Singapore, 4 Engineering Drive 3, 117576, Singapore

[c] Hysitron, Inc., 10025 Valley View Rd, Minneapolis, MN 55344, USA

[d] Department of Physics, National University of Singapore, 2 Science Drive 3, 117542, Singapore

## Abstract

To determine the friction coefficient of graphene, micro-scale scratch tests are conducted on exfoliated and epitaxial graphene at ambient conditions. The experimental results show that the monolayer, bilayer, and trilayer graphene all yield friction coefficients of approximately 0.03. The friction coefficient of pristine graphene is less than that of disordered graphene, which is treated by oxygen plasma. Ramping force scratch tests are performed on graphene with various numbers of layers to determine the normal load required for the probe to penetrate graphene. A very low friction coefficient and also its high pressure resistance make graphene a promising material for antiwear coatings.

Graphene, a one-atom-thick planar sheet of carbon atoms, has been studied intensively in the last few years due to its unique characteristics. However, only a few studies investigate its mechanical properties [1, 2]. In particular, micro-scale friction coefficient of graphene has never been investigated despite its promising potential for low-friction antiwear coatings. In this report, scratch tests are conducted to determine the friction coefficient of mechanically exfoliated graphene on $SiO_2$ and epitaxial graphene on SiC under ambient conditions. Although the electrical properties of graphene are sensitive to the number of layers, no thickness dependence of friction coefficient is observed. The friction coefficient measurements on disordered graphene treated by oxygen plasma show that disorder in graphene increases the friction coefficient. Ramping force scratch tests are performed on graphene samples with different numbers of layers to determine the normal load required for the probe to penetrate through the graphene, inducing failure of the film. This load is referred to here the critical load.

[*] corresponding author. Tel.:+65 65167217. Fax: +65 67791103. E-mail address: eleyang@nus.edu.sg (H. Yang).

Mechanically cleaved graphene is transferred to a highly p-doped Si substrate covered by a 300 nm thick $SiO_2$ layer [3]. Epitaxial graphene on SiC is prepared by annealing chemically etched n-type Si-terminated 6*H*-SiC (0001) at 850 ºC under silicon flux for 2 minutes in ultrahigh vacuum [4]. Number of layers and quality of exfoliated graphene and epitaxial graphene on SiC are determined by Raman spectroscopy [5]. A TI 950 TriboIndenter™ (Hysitron, Inc.) instrument is employed to perform scratch tests on graphene using a diamond 90º conical probe with a 1 μm radius of curvature. Scratches are 2 μm long in 60 seconds.

Single-, bi-, and tri-layer exfoliated graphene samples are identified by Raman spectra as shown in Fig. 1(a). In-plane vibrational *G* band (1580 $cm^{-1}$) and two-phonon 2*D* band (2670 $cm^{-1}$) are clearly dominant without any indication of disorder *D* band and the number of layers of the graphene samples are determined by estimating the width of 2*D* peak [5]. Epitaxial graphene gives an additional Raman peak at 1513 $cm^{-1}$ due to the scattering from the SiC substrate [4]. Fig. 1(b) shows the optical image of graphene flake derived from mechanical exfoliation. Fig. 1(c) and (d) show error signal and topographical *in-situ* scanning probe microscopy (SPM) images after scratch tests, respectively. Although the magnitude of the error signal depends on the feedback parameters of the scan, the error signal image, the difference between the actual force and the set point at any given moment, is useful because it often shows more contrast than the accompanying topography image. Deformed parts of graphene after scratch tests are marked by red circles.

Fig. 2(a) shows a plot of normal force and lateral force versus time (*T*) from the scratch test on single layer graphene. For plotting purposes, time scales are offset so that $T$=0 corresponds to the beginning of the main portion of the scratch. After performing similar tests on bi- and tri-layer graphene, the friction coefficients are calculated in Fig. 2(b). The results show that all the graphene samples with different number of layers yield similar friction values, 0.03. Filleter *et al.* reported a wide range of friction coefficient from 0.004 to 0.07 in single and bilayer epitaxial graphene films [2]. The difference from our data can be attributed to measurement environments [6]. First of all, our experiment is conducted at ambient conditions and their experiment is under ultra high vacuum conditions. It is well known that the existence of water can influence the friction coefficient, therefore it is reasonable to have a different friction coefficient depending on the measurement conditions [6]. The different value could be also attributed to the difference in the size of probes; Filleter used an atomic force microscopy (AFM) based system, whereas we used a larger diamond probe with a 1 µm radius. This is in line with a previous study, where a large difference in the friction coefficient of graphite in atomic scale and its macroscopic value was reported from a highly oriented pyrolytic graphite surface [7]. Lee *et al*. reported the thickness dependence of frictional characteristics of graphene [1]. The different results from our experiments can be also explained by the different size of probe. Our probe radius was 1 µm whereas the AFM experiment by Lee used a ~5 nm probe. The adhesion-induced puckering mechanism may behave differently with larger contacts. With our probe, which was 200 times blunter, the 1 to 3 layer graphene could be equally able to bend around the gradual curvature of our probe comparing to a few nanometer probe.

Frictional characteristic between pristine graphene and disordered graphene is also compared. Damage on graphene is intentionally introduced by oxygen plasma treatment. The level of damage in exfoliated and epitaxial graphene is measured by Raman spectra as shown in the insets of Fig. 2 (c) and (d), respectively. A pronounced D band peak in the insets of Fig. 2 (c) and (d) indicates that both samples are damaged after the plasma treatment. It is found that the friction coefficient of both exfoliated graphene and epitaxial graphene increases after the plasma treatment. Physically, a surface dislocation, vacancies, or corrugation created by oxygen plasma treatment can be the origin of the disorder [8]. These components can possibly contribute to increase the polarity of graphene surface [3]. The defects convert pristine graphene to be more adhesive, and the increase in van der Waals force between the probe tip and graphene layer can work as a drag force when the tip moves laterally. This phenomenon results in an increase in the friction coefficient of graphene in the presence of defects.

Ramping force scratch tests are performed on exfoliated and pristine graphene samples with different numbers of layers to determine the critical load corresponding to film failure in Fig. 3. Failure events may be found where the probe produces crack, fracture, or breakthrough at the interface between graphene and $SiO_2$. The failure events of the film are normally symbolized by a combination of sudden changes in the lateral force data as can be seen in Fig. 3(a-c). The critical load is defined as the normal force applied to the scratch probe at the time when failure is detected. The vertical blue lines in Fig. 3 indicate the location of the critical points. Various loads are used on different samples and critical points are observed at different normal forces. For example, the measured critical loads for different single layer graphene range from 450 to 2250 μN. The estimated values of mean pressure at the critical failure point can be calculated by dividing the applied force, *P*, by the estimated contact area, *A* at the time of failure. *P* is a specified test parameter while *A* can be estimated through a probe calibration that relates contact area to depth obtained on a fused quartz standard sample. In this case, a fused quartz sample with a reduced modulus of 69.6 GPa was utilized. Through the above method, the mean pressure was estimated to be 6.78 GPa when the critical load is 1400 μN, the displacement is 29.2 nm, and the contact area is $2.07 \times 10^5$ $nm^2$. Note that the above mean pressure is in a concentrated contact where pressure varies greatly with position.

The results of critical load test on graphene indicate that the critical load can vary significantly between graphene samples, even with the same number of layers. Since the correlation between film strength and the critical load is complicated and influenced by many factors such as the roughness of $SiO_2$, the surface energy of $SiO_2$, and the scratch testing parameters, further studies are required to understand the relationship between film strength and the critical load. One notable observation is that all graphene samples yield a similar friction coefficient of 0.03 prior to the failure load.

When normal force is applied for scratch tests, the normal displacement of the probe is monitored simultaneously as plotted in Fig. 3 (d). From the critical load measurement, it is confirmed that graphene itself is not delaminated or peeled off from the substrate prior to the failure point. Rather graphene is bent and displaced more than 50 nm together with the supporting substrate when it reaches the failure point in Fig. 3(d) as the substrate underneath the graphene is deformed under the applied load. One atom-thick-layer material holds the normal displacement for more than 100 times of its thickness. Graphene can be employed for antiwear coatings with its very low friction coefficient and elastic characteristic. In particular, graphene can be an ideal material as lubricant layer for the next generation magnetic media of hard disks since future tribology technology of hard disks requires sub-2 nm of the disk overcoat.

We investigate the friction coefficient of graphene as a function of number of layers at ambient conditions. The friction coefficient (0.03) of graphene is independent of the thickness of graphene. The friction coefficient of pristine graphene and disordered graphene is compared and it is found that defects in graphene increase the friction coefficient. Ramping force scratch tests show that graphene is very elastic, maintaining its integrity during scratch tests with a normal displacement of more than 50 nm. Graphene can be engineered as an antiwear coating material because of a very low friction coefficient and high pressure resistance at ambient conditions.

**Acknowledgement**

This work is supported by the Singapore National Research Foundation under CRP Award No. NRF-CRP 4-2008-06.

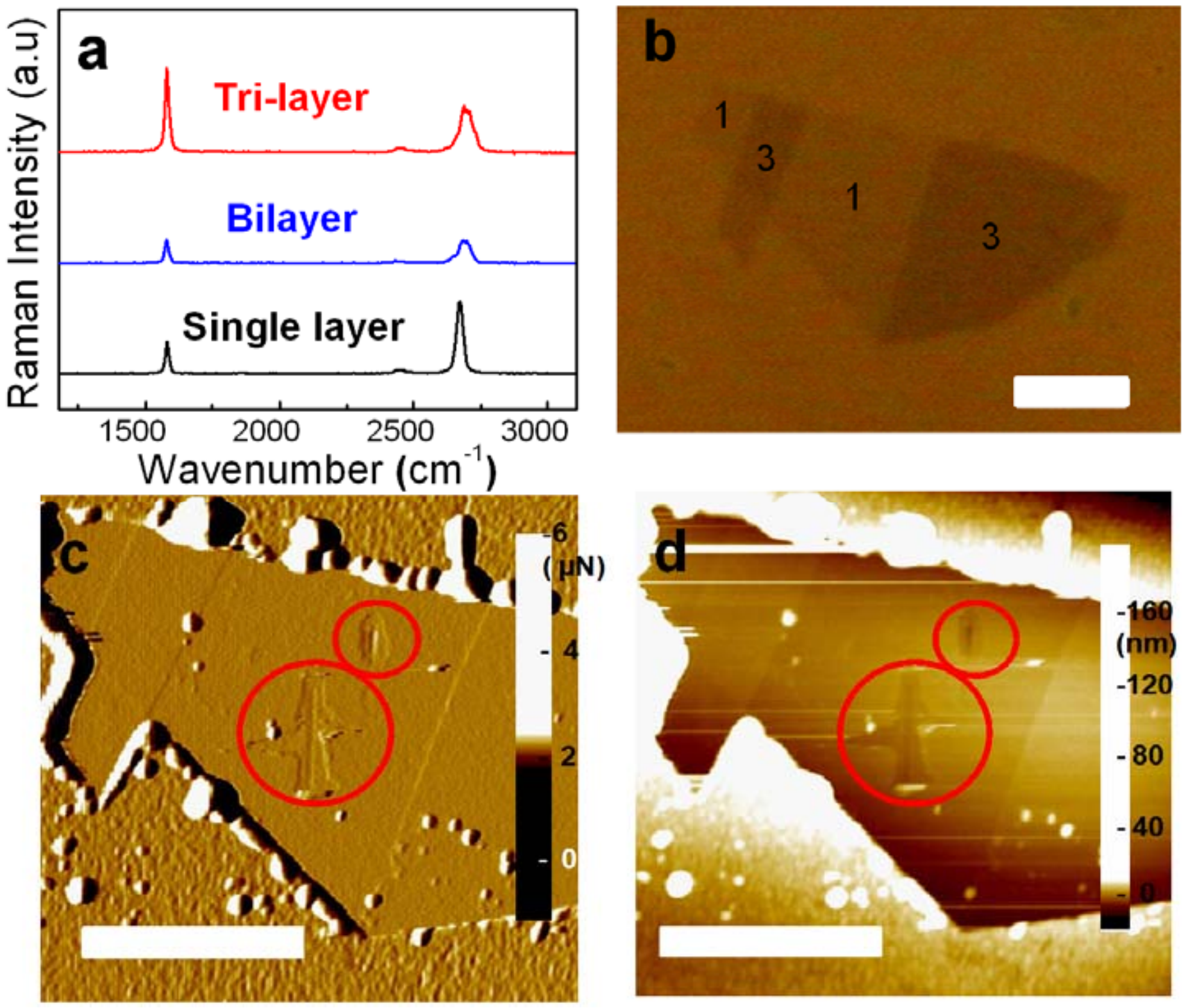

Fig. 1 - (a) Raman spectra of single, bi-, and tri-layer graphene. (b) Optical image of mechanically exfoliated graphene on $SiO_2$. The number in (b) indicates the number of layer. Error signal (c) and topographical (d) *in-situ* SPM images in the contact mode. Deformed parts of graphene after scratch tests are marked by red circles in (c,d). All scale bars are 3 µm.

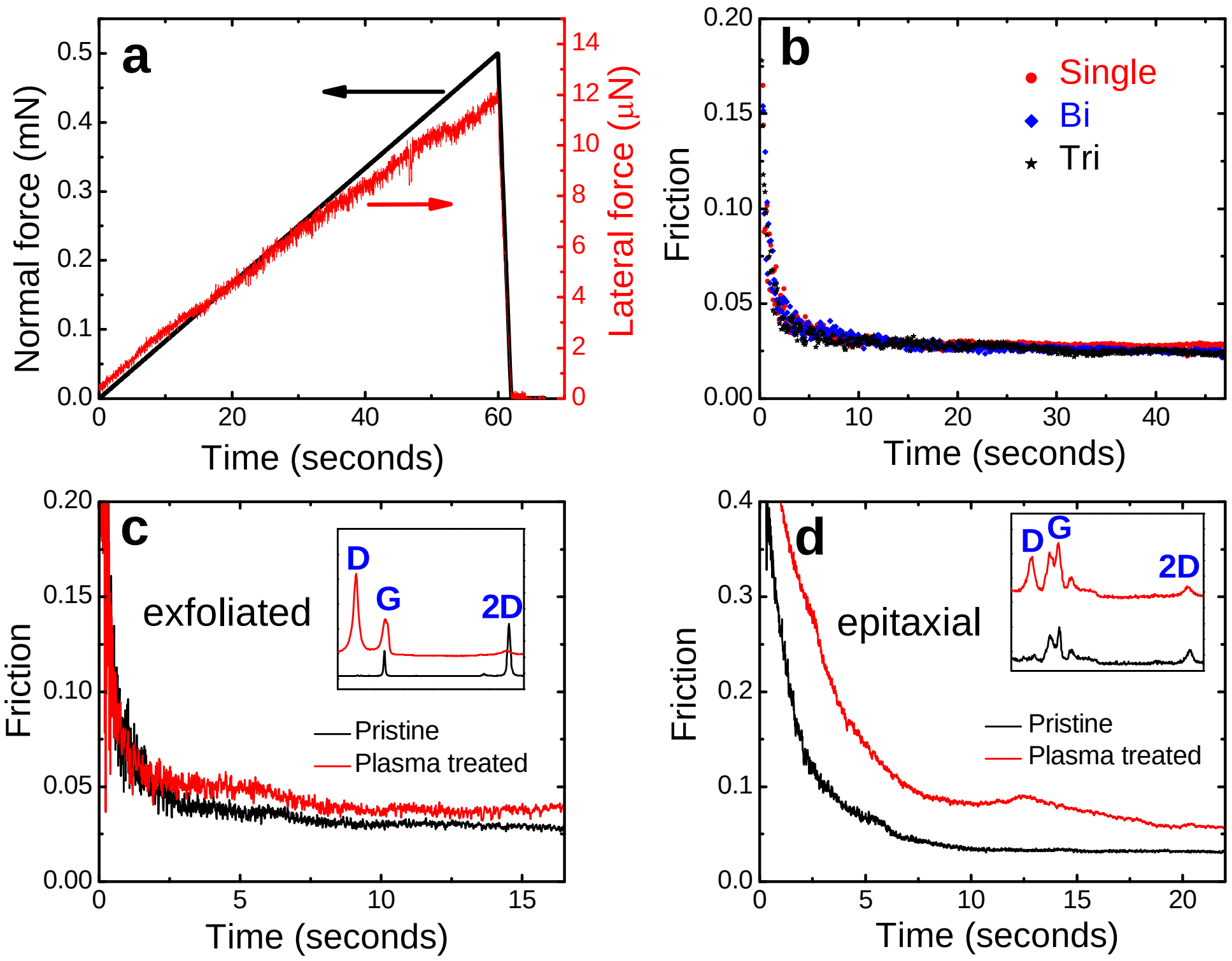

Fig. 2 - (a) Normal force and lateral force on graphene versus time. (b) Friction coefficient (lateral force/normal force) versus time of single, bi-, and tri-layer graphene. (c) Friction coefficient versus time of mechanically exfoliated graphene without and with defects. The black solid lines represent the friction and Raman data before oxygen plasma treatment, and the red lines are for the samples after oxygen plasma treatment. The inset in (c) is Raman spectra of mechanically exfoliated graphene without and with defects. (d) Friction coefficient versus time of epitaxial graphene without and with defects. The inset in (d) is Raman spectra of epitaxial graphene without and with defects.

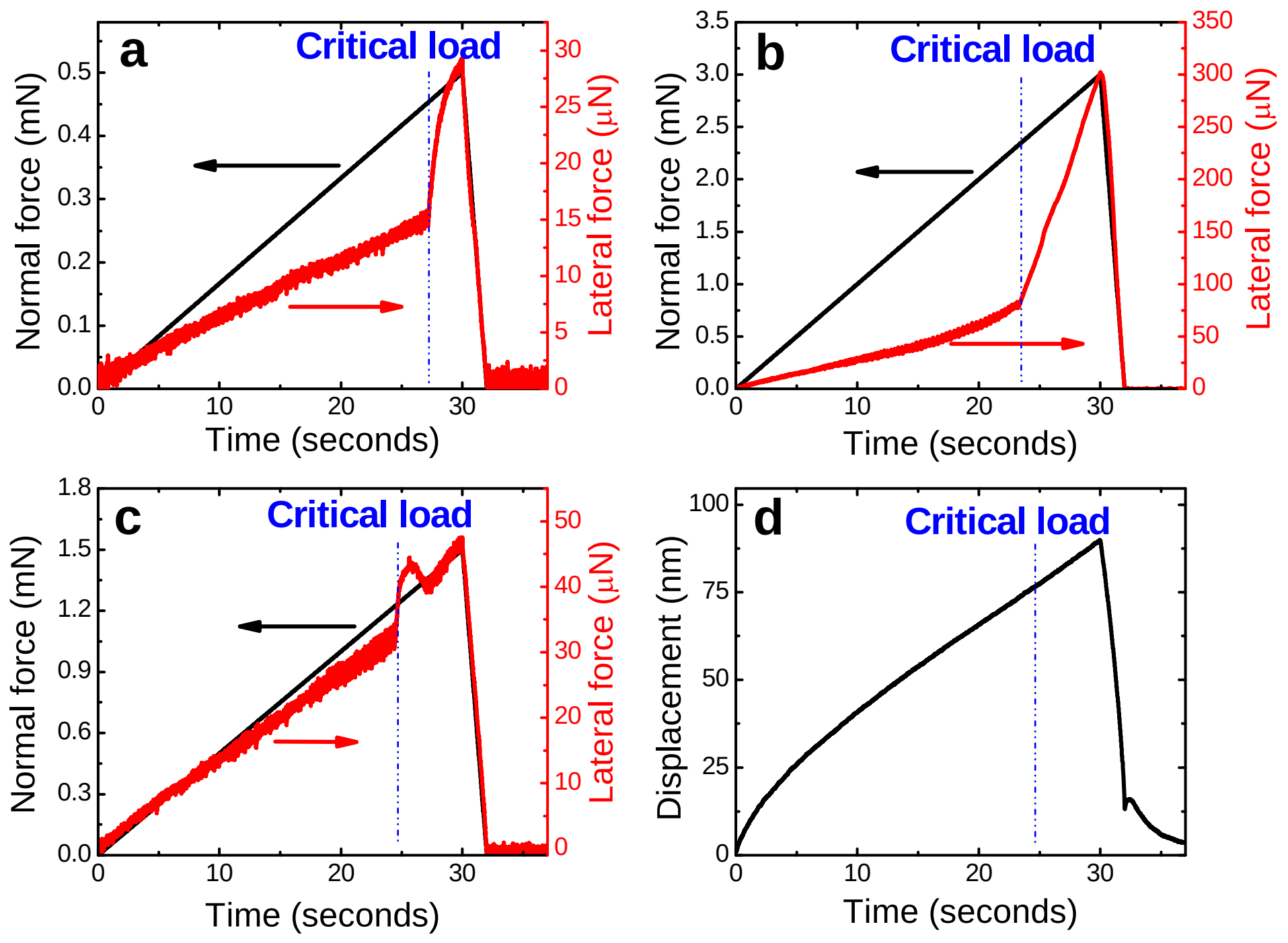

Fig. 3 - Normal force and lateral force versus time on single layer (a, b) and bilayer (c) graphene. (d) Normal displacement of probe versus time of the sample in (c). All of the tested samples are exfoliated and pristine graphene.